\documentclass[sigconf,natbib,screen=true,review=false,anonymous=false]{acmart}

\acmSubmissionID{5292}

\usepackage{amsmath}
\usepackage{amsfonts}
\usepackage{dsfont}
\usepackage{mathtools}
\usepackage{graphicx}
\usepackage{booktabs} 
\usepackage{numprint}
\usepackage{hyperref}
\npthousandsep{,}
\npdecimalsign{.}
\usepackage[inline]{enumitem}
\usepackage{balance}
\usepackage[skip=0pt]{caption}
\usepackage{multirow}
\usepackage{siunitx}
\usepackage{acronym}
\usepackage{subfig}
\usepackage{xcolor}
\usepackage{colortbl}
\usepackage{tabularx}
\usepackage{etoolbox,siunitx}
\usepackage{caption}
\usepackage{adjustbox}
\usepackage{algorithm}
\usepackage{algpseudocode}
\usepackage{xspace}
\robustify\bfseries

\copyrightyear{2023}
\acmYear{2023}
\setcopyright{rightsretained}
\acmConference[SIGIR '23]{Proceedings of the 46th International ACM SIGIR Conference on Research and Development in Information Retrieval}{July 23--27, 2023}{Taipei, Taiwan}
\acmBooktitle{Proceedings of the 46th International ACM SIGIR Conference on Research and Development in Information Retrieval (SIGIR '23), July 23--27, 2023, Taipei, Taiwan}
\acmDOI{}
\acmISBN{}

\definecolor{darkgreen}{rgb}{0.0, 0.2, 0.13}
\definecolor{asparagus}{rgb}{0.53, 0.66, 0.42}
\definecolor{cambridgeblue}{rgb}{0.64, 0.76, 0.68}

\newcommand{\header}[1]{\vspace{1mm}\noindent\textbf{#1}.}

 \algrenewcommand\algorithmicrequire{\textbf{Input:}}
\algrenewcommand\algorithmicensure{\textbf{Output:}}

\AtBeginDocument{%
  \providecommand\BibTeX{{%
    \normalfont B\kern-0.5em{\scshape i\kern-0.25em b}\kern-0.8em\TeX}}}

\newtheorem{corollary }{Corollary}

\theoremstyle{remark}
\newtheorem*{remark}{Remark}

\newcommand{\RomanNumeralCaps}[1]{\MakeUppercase{\romannumeral #1}}
    
\newcommand{\noOutlier}{normal\xspace}
\newcommand{\withOutlier}{abnormal\xspace}
\newcommand{\price}{price\xspace}
\newcommand{\promotion}{promotion\xspace}
\newcommand{\discount}{discount\xspace}
\newcommand{\stockQuantity}{stock quantity\xspace}
\newcommand{\rating}{rating\xspace}
\newcommand{\select}{select\xspace}
\newcommand{\titleLength}{title length\xspace}

\newcommand{\ourmodel}{OPBM\xspace}
\newcommand{\bolOpbm}{OPBM$_{Real}$\xspace}
\newcommand{\normalOpbm}{OPBM$_\mathcal{G}$\xspace}
\newcommand{\mixtureModel}{OPBM$_{Mixture}$\xspace}
\newcommand{\target}{target\xspace}
\newcommand{\condone}{condition \RomanNumeralCaps{1}\xspace}
\newcommand{\condtwo}{condition \RomanNumeralCaps{2}\xspace}
\newcommand{\ourmodelsimple}{\ourmodel$_{lazy}$\xspace}
\newcommand{\propdata}{proprietary\xspace}

\newcommand{\yahoo}{Yahoo!\xspace}
\newcommand{\mslr}{MSLR\xspace}

\acrodef{CTR}{click-through rate\xspace}
\acrodef{SERP}{search engine Result Page\xspace}
\acrodef{LTR}{learning to rank\xspace}
\acrodef{CLTR}{counterfactual learning to rank\xspace}
\acrodef{ULTR}{unbiased learning to rank\xspace}
\acrodef{NDCG}{normalized discounted cumulative gain\xspace}
\acrodef{CE}{cross entropy\xspace}
\acrodef{PBM}{position-based model\xspace}
\acrodef{OPBM}{outlier-aware position-based model\xspace}
\acrodef{EM}{standard expectation maximization\xspace}
\acrodef{REM}{regression-based expectation maximization\xspace}
\acrodef{GBDT}{gradient boosted decision tree\xspace}
\acrodef{normal-OPBM}{gaussian distribution-based outlier-aware position-based model\xspace}
\acrodef{IPS}{inverse propensity scoring\xspace}

\allowdisplaybreaks

\author{Fatemeh Sarvi}
\orcid{0000-0002-6284-1636}
\affiliation{%
  \institution{
  AIRLab, University of Amsterdam
  \city{Amsterdam}
  \country{The Netherlands}%
  }
}  
\email{f.sarvi@uva.nl}

\author{Ali Vardasbi}
\orcid{0000-0002-9342-5272}
\affiliation{%
  \institution{University of Amsterdam}
  \city{Amsterdam}
  \country{The Netherlands}
}
\email{a.vardasbi@uva.nl}

\author{%
Mohammad Aliannejadi
}
\orcid{0000-0002-9447-4172}
\affiliation{%
  \institution{
  University of Amsterdam
  \city{Amsterdam}
  \country{The Netherlands}%
  }
}  
\email{m.aliannejadi@uva.nl}

\author{Sebastian Schelter}
\orcid{0000-0003-4722-5840}
\affiliation{%
  \institution{
  University of Amsterdam
  \city{Amsterdam}
  \country{The Netherlands}%
  }
}
\email{s.schelter@uva.nl}

\author{Maarten de Rijke}
\orcid{0000-0002-1086-0202}
\affiliation{%
  \institution{
  University of Amsterdam
  \city{Amsterdam}
  \country{The Netherlands}%
  }  
}  
\email{m.derijke@uva.nl}

\copyrightyear{2023} 
\acmYear{2023} 
\setcopyright{rightsretained} 
\acmConference[SIGIR '23]{Proceedings of the 46th International ACM SIGIR Conference on Research and Development in Information Retrieval}{July 23--27, 2023}{Taipei, Taiwan}
\acmBooktitle{Proceedings of the 46th International ACM SIGIR Conference on Research and Development in Information Retrieval (SIGIR '23), July 23--27, 2023, Taipei, Taiwan}\acmDOI{10.1145/3539618.3591745}
\acmISBN{978-1-4503-9408-6/23/07}

\begin{document}

\title[On the Impact of Outlier Bias on User Clicks]{On the Impact of Outlier Bias on User Clicks}

\renewcommand{\shortauthors}{Sarvi et al.}

\begin{abstract}
User interaction data is an important source of supervision in \ac{CLTR}.
Such data suffers from presentation bias.
Much work in \ac{ULTR} focuses on position bias, i.e., items at higher ranks are more likely to be examined and clicked.
Inter-item dependencies also influence examination probabilities, with \emph{outlier} items in a ranking as an important example. 
Outliers are defined as items that observably deviate from the rest and therefore stand out in the ranking. 
In this paper, we identify and introduce the bias brought about by outlier items: users tend to click more on outlier items and their close neighbors.

To this end, we first conduct a controlled experiment to study the effect of outliers on user clicks. 
Next, to examine whether the findings from our controlled experiment generalize to naturalistic situations, we explore real-world click logs from an e-commerce platform. 
We show that, in both scenarios, users tend to click significantly more on outlier items than on non-outlier items in the same rankings. 
We show that this tendency holds for all positions, i.e.,~for any specific position, an item receives more interactions when presented as an outlier as opposed to a non-outlier item.  
We conclude from our analysis that the effect of outliers on clicks is a type of bias that should be addressed in \ac{ULTR}. 
We therefore propose an outlier-aware click model that accounts for both outlier and position bias, called \ac{OPBM}. 
We estimate click propensities based on \ac{OPBM}; through extensive experiments performed on both real-world e-commerce data and semi-synthetic data, we verify the effectiveness of our outlier-aware click model.
Our results show the superiority of OPBM against baselines in terms of ranking performance and true relevance estimation.
\end{abstract}

\ccsdesc[300]{Information systems~Learning to rank}
\keywords{Click bias; Outliers; Counterfactual learning to rank}

\maketitle
\acresetall


\section{Introduction}
\label{sec:intro}
Ranking systems optimize ranking decisions to increase user satisfaction. 
Implicit user feedback is an important source of supervision that reflects the preferences of actual users.
However, user interaction data (e.g., clicks) suffers from presentation bias, which can make its na\"ive use as training data highly misleading~\citep{joachims2007evaluating}.

Much work in \ac{ULTR} focuses on position bias~\citep{agarwal2019estimating,joachims2005accurately,joachims2017unbiased,wang2018position}, i.e., the phenomenon that higher-ranked results are more likely to be examined and thus clicked by users~\citep{joachims2005accurately} than lower-ranked results. 
Besides position there are several other factors that affect users' examination model and clicks~\citep{sarvi2022understanding, wu2021unbiased,ovaisi2020correcting,abdollahpouri2017controlling,chen2019correcting}. 
Previous work has shown that inter-item dependencies can influence user judgments of relevance and the examination order of items~\citep{wu2021unbiased,chuklin-2015-click,sarvi2022understanding}.
The existence of \emph{outlier} items is a specific case of inter-item dependencies~\citep{sarvi2022understanding}.
\citet{sarvi2022understanding} define outliers in a ranking as items that observably deviate from the rest of the list w.r.t.\ item features, such that they stand out and catch users' attention.
For instance, in an e-commerce search scenario, if only one item on the page features a ``Best Seller'' tag, it can be considered as an outlier, because the tag differentiates it from the rest of the items in the ranking, thereby attracting users' attention.

\header{Outlier bias}
An outlier in a list of items can alter the examination probabilities, such that the probability of examination is higher for the outlier item (if it exists) and its neighboring items than the probability assigned by the position bias assumption~\citep{sarvi2022understanding}.

Although it has been shown that outliers affect  examination probabilities~\cite{sarvi2022understanding}, their impact on user click behavior is unknown. 
In this work, we hypothesize that clicks are biased by the existence of outliers.
We refer to this phenomenon as \emph{outlier bias} and aim to understand and address this effect. 
To begin, we conduct a user study where we compare the \ac{CTR} for specific items in two conditions: once shown as outliers and once as non-outlier items in the list. 
We find that users behave differently in relation to an item given its outlierness condition. 
The \ac{CTR} of a specific item is consistently higher when it is presented as an outlier item than when it is a non-outlier item in a ranking.
Next, to examine whether these findings can be generalized to naturalistic situations we perform an analysis on real-world search logs from \href{http://www.bol.com}{Bol.com}, a popular Europe-based e-commerce platform. The results confirm the findings of our user study. In addition, we observe that, on average, outlier items receive significantly more clicks than non-outlier items in the same lists. Moreover, users tend to interact more with lists that contain at least one outlier.

\header{Outlier bias vs.\ context bias}
We find that outlier bias affects user clicks such that users are more likely to interact with items that are presented as outliers, as well as their neighboring items.
The closest concept to outlier bias is context bias in news-feed recommendation~\citep{wu2021unbiased}.
In the presence of context bias \ac{CTR} is lower for items when surrounded by at least one very similar item than when they are surrounded by non-similar items. 
This is different from outlier bias, which emphasizes the \emph{difference} between the outlier and the rest of the list. 
Moreover, observability is a key factor in detecting outliers in ranking as defined by~\cite{sarvi2022understanding}; this is not the case in context bias.

\header{Accounting for outliers}
Based on the findings of our user study and log analysis, we conclude that one should account for the effect of outliers when unbiasing user clicks for \ac{ULTR}.
To this end, we propose a click model, based on the examination hypothesis, called \acfi{OPBM}, which accounts for both outlier and position bias. 
\ac{OPBM} assumes the probability of a click depends on 
\begin{enumerate*}[label=(\roman*)]
    \item examination,
    \item relevance, and
    \item the outlier's position (if it exists)
\end{enumerate*}. 
We use \acl{REM} to estimate the click propensities based on our proposed click model, \ac{OPBM}. We verify the effectiveness of our outlier-aware model for estimating propensities in the presence of both position bias and outlier bias. 
Following~\citep{ai2018unbiased, joachims2017unbiased,oosterhuis2020policy} we use a semi-synthetic setup for the experiments; the true relevance labels provided in this setup  allows for evaluating the relevance estimation. Furthermore, using simulated clicks we are able to control the severity of position bias and outlier bias. 
The results of our experiments show the superiority of \ac{OPBM} against baselines in terms of ranking performance (NDCG@10) and true relevance estimation. 

\header{Main contributions}
The main contributions of this work are: 
\begin{enumerate*}[label=(\roman*)]
    \item we identify and study a new type of click bias, originating from inter-item dependencies, called outlier bias;
    \item through extensive analyses of both user study results and real-world search logs, we confirm our hypothesis about the existence of outlier bias;
    \item to address this effect we propose an outlier-aware click model that accounts for outlier items (if they exist), as well as position bias;
    \item using an empirical analysis based on real-world data and semi-synthetic experiments we show the effectiveness of our outlier-aware model in estimating click propensities; and
    \item we make the data from our user study plus the code that implements our baselines and \ac{OPBM} publicly available.
\end{enumerate*}
\section{Outliers in ranking}
\label{sec:background}

Outliers in ranking are items that observably stand out among the window of items that are presented to a user at once. We use the following definitions from~\citep{sarvi2022understanding} to introduce so-called outliers: 

\begin{definition}[Observable feature]
An observable item feature, $\mathcal{F}$, is a characteristic of an item in a list that can be purely presentational in nature (e.g., image, title font size, and discount tag).
\end{definition}

\begin{definition}[Degree of outlierness]
Let $\mathcal{M}$ be any outlier detection method, and $\mathcal{F}_i$ an observable feature corresponding to item $i$, in the context of all items in the list, $\mathcal{C}$.
The degree of outlierness for item $i$ is the value calculated by $\mathcal{M}$ for $\mathcal{F}_i$ w.r.t.~$\mathcal{C}$ shown as $\mathcal{M}(\mathcal{F}_i|\mathcal{C})$. This value indicates how much the corresponding item differs from the other elements of the set w.r.t.~$\mathcal{F}$.
\end{definition}

\begin{definition}[Outliers in ranking]
Let $\mathcal{M}$ be any outlier detection method;
we call item $i$ in a ranked list an outlier, if $\mathcal{M}$ identifies it as an outlier w.r.t.~an observable feature, $\mathcal{F}$, based on the degree of outlierness, and in the context of the list.
\end{definition}

\noindent%
In Section~\ref{sec:outlier-impact:logs} we describe our choices of observable features and outlier detection method used in this paper.

\section{Impact of Item Outlierness on Clicks}
\label{sec:outlier-impact}

\citet{sarvi2022understanding} show that outlier items receive more attention from users. 
However, it is not known whether an item's outlierness affects users' clicks as well. In this section we answer our first research question: \begin{enumerate*}[nosep,label=(\textbf{RQ\arabic*}),leftmargin=*]
    \item does outlier bias exist in rankings of items?\label{RQ1} 
\end{enumerate*}
To this end we first conduct a user study to examine the outlierness effect as the only variable factor influencing the clicks. 
Next, we need to examine whether the findings of our study can be generalized to naturalistic situations. 
In other words, we seek to establish ecological validity~\citep{lewkowicz2001concept, andrade2018internal}. 
To this end, in Section~\ref{sec:outlier-impact:logs} we explore real-world click logs to confirm our findings. 

\subsection{User study}
\label{sec:outlier-impact:user-study}
In this section, we present the results of our user study. Our main goal is to learn whether the outlierness of an item affects user clicks, independent of the item's relevance and position.

\header{Setup} We mimic \href{http://www.bol.com}{Bol.com}, a popular European online marketplace.
We ask participants to interact with search engine result pages as they normally would, and find items they prefer and think are relevant. We focus on a list view, with 20 items on each page, and participants are able to scroll the list to see all items. 
We have two queries; for each query, we show one specific item once as an outlier and once as a regular item. We call this specific item the \emph{\target}, and these two variant presentations \emph{\condone} and \emph{\condtwo}, respectively. 
In \condone the \target is an outlier w.r.t.~a set of observable features, such as item category,\footnote{Note that this feature can affect the outlierness w.r.t.~the item's image as well.} price, discount tag, and star rating.
We aim to compare users' behavior between these two conditions for each query. We keep other factors such as relevance and position bias unchanged between the conditions.
To eliminate the effect of position bias we always show the item at rank 4, and to maintain the same degree of relevance to the query we only change the surrounding items to change the outlierness of the \target item.\footnote{To examine our hypothesis about an inter-item dependency, here we assume that the relevance of a document is only dependent on the query.}

We also have a Qualtrics~\citep{qualtrics} survey.
It contains the task instruction, multiple choice questions about the instructions, queries, and links to the examples, and a few demographic questions at the end.
In the instructions, we describe the overall goal of the research and ask participants to read the instructions carefully.
We describe what it means to interact with a result page in terms of exploring the results, scrolling the list, and clicking on items that seem interesting.
Participants can click on an item to open the item's detail page. 
In our instructions we encourage participants to click on items they find interesting, however, clicking is not mandatory.
We instruct participants to first read and understand the query, and then scan the result page as if they submitted the query themselves.

\header{Participants} We recruit 40 workers, based in countries where our marketplace is active, from the Prolific platform~\citep{prolific}. From the participants, 14 are female, 23 are male, and 3 listed other genders. The majority of participants (27) are between 25 and 44 years old, with 10 participants younger and 3 older; 33 participants reported that they shop online at least once a month.

\header{Metrics} For reporting we consider three measures based on participants' interactions with rankings: \begin{enumerate*}[label=(\roman*)]
    \item revisit count, which indicates how many times on average participants viewed an item (due to scrolling),
    \item mouse hover time that shows the amount of time on average participants spent on an item, and
    \item \ac{CTR} for the \target item in each condition, which is our main metric in this study.
\end{enumerate*}   

\header{Findings}
We expect to see more interactions with the \target in \condone. Since we keep other factors unchanged between the two conditions, we can attribute any difference in user behavior to the inter-item dependencies.

\begin{figure}
    \begin{tabular}{@{}c@{~}c@{~}c@{}}
    \subfloat[Revisit count]{%
        \includegraphics[,height=3.64cm,width=.5\linewidth]{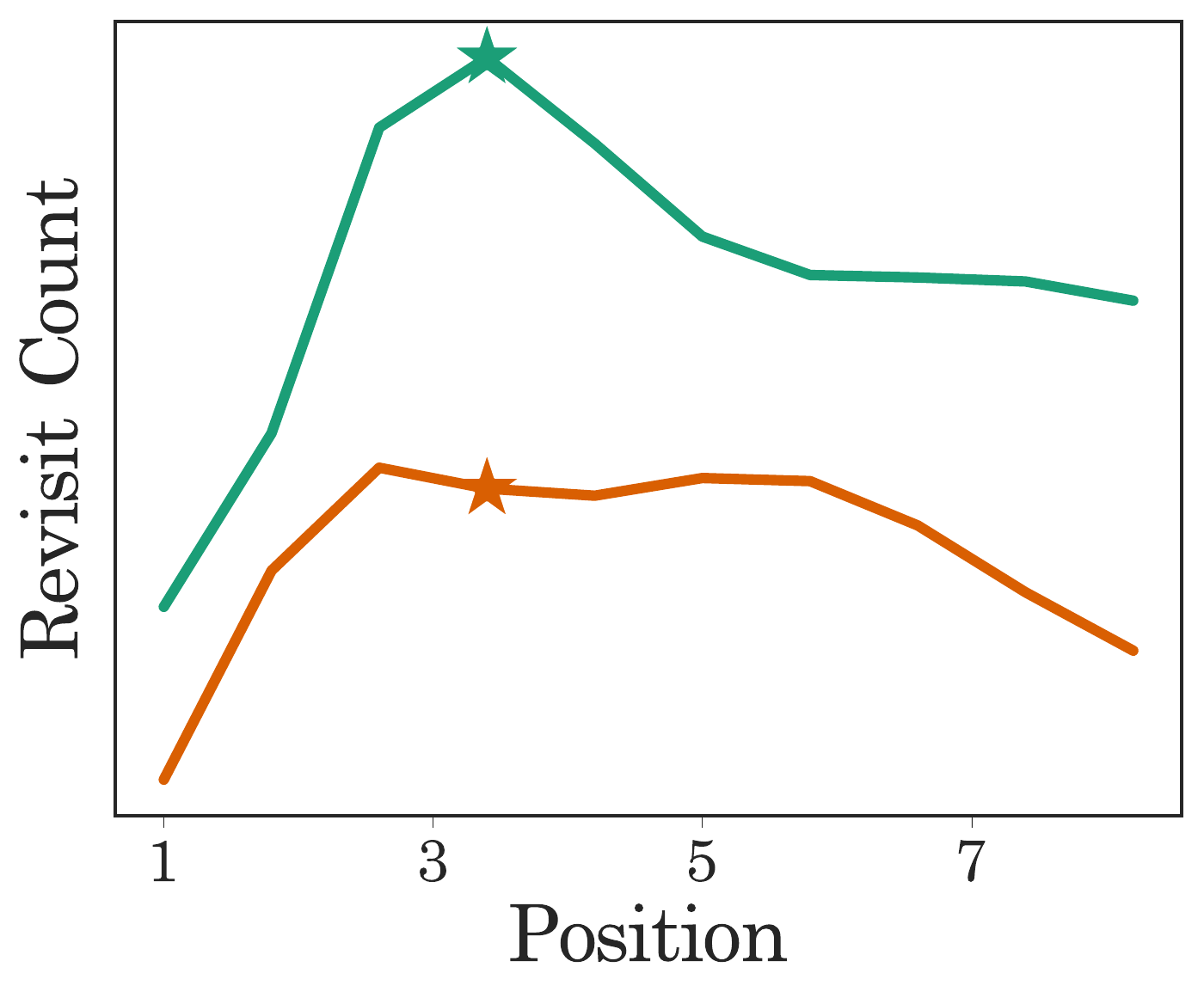}
          \label{fig:user-study:revits}}
     &
    \subfloat[Mouse hovering time]{%
          \includegraphics[height=3.64cm,width=.5\linewidth]{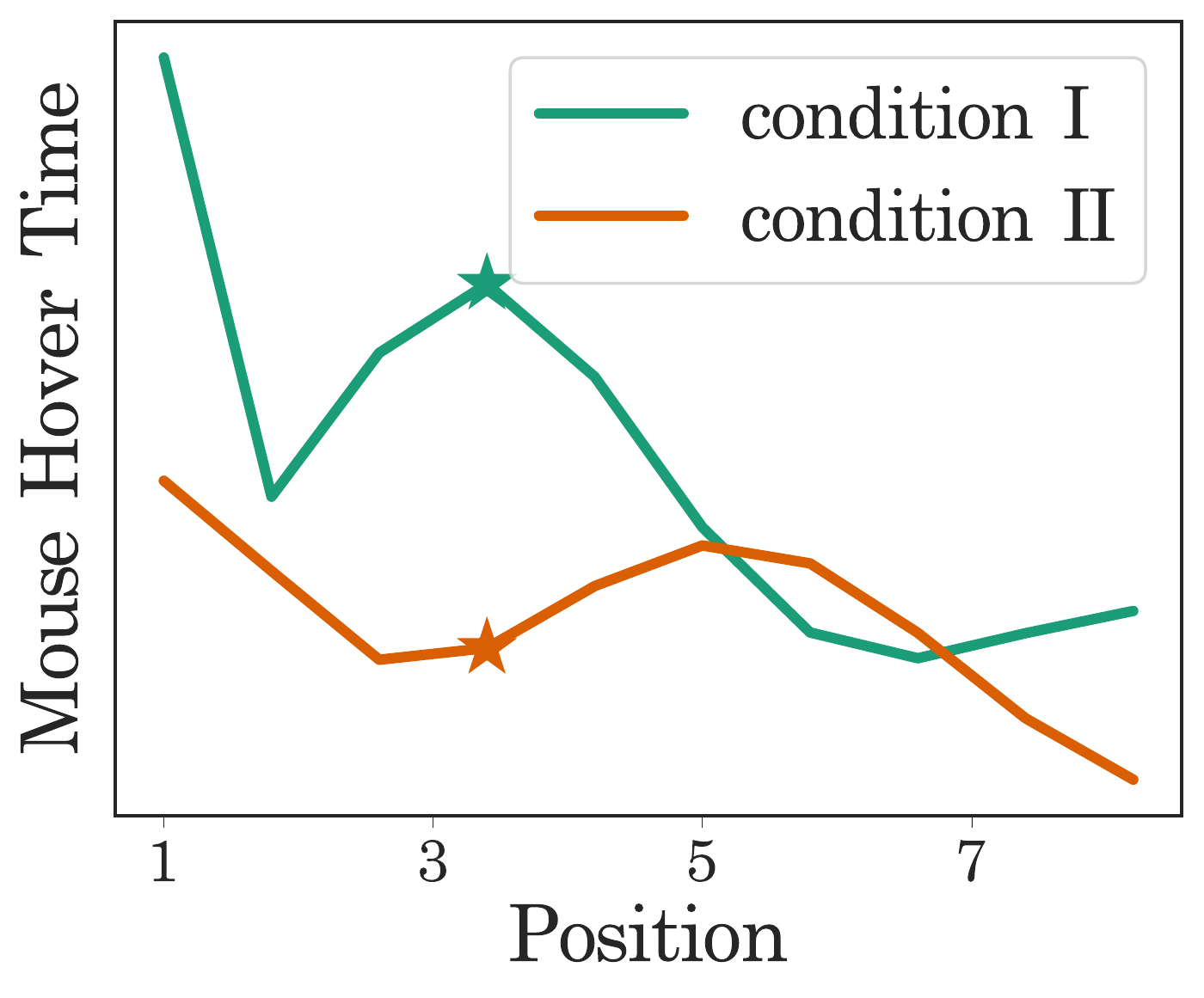}
          \label{fig:user-study:hover}} 
    \end{tabular}
\caption{Revisit count (a) and mouse hovering time (b) for the two conditions of one of our user study examples. The position of the \target item is marked by an asterisk. The plots show that the user engagement with the \target item is higher when presented as an outlier (\condone)}  
\label{fig:user-study}
\end{figure}

Figure~\ref{fig:user-study} depicts the revisit counts (Figure~\ref{fig:user-study:revits}) and mouse hovering time (Figure~\ref{fig:user-study:hover}) for different positions and conditions of one example. Both plots show that the user engagement with the \target item is higher when it is presented as an outlier. We see the same pattern in the second example. On average, participants revisited the outlier item more often and spent more time examining it. These findings are in line with the results of the eye-tracking experiments conducted by~\citet{sarvi2022understanding}, which suggest that, on average, outlier items receive more attention from users. However, our main goal is to study if this increased attention leads to more clicks.

Table~\ref{table:user-study} reports the \ac{CTR} for the \target item in both examples and for the two conditions. In both examples we see a large difference between the \ac{CTR} reported for the different conditions, suggesting that when the \target item is shown as an outlier it receives more clicks as well as more exposure.  

\begin{table}[t]
   \caption{\ac{CTR} of the \target item's position in both examples of our user study. The \target item recieves more clicks when shown as an outlier (\condone).}
   \label{table:user-study}
   \centering
   \begin{tabular}{lcc}
     \toprule
        &  \condone  &  \condtwo \\
      \midrule
      Query 1 & 0.944 &  0.166 \\
      Query 2 & 0.880 & 0.091 \\
      \bottomrule
   \end{tabular}
\end{table}

\subsection{Real-world click logs}
\label{sec:outlier-impact:logs}

The findings of Section~\ref{sec:outlier-impact:user-study} confirm, in a controlled experimental setup, that an item's outlierness can influence users' click behavior. However, we still need to examine the ecological validity of this hypothesis.
To this end, we present our observations of click exploration of real-world search logs from our e-commerce platform. We are specifically interested in exploring the data to study the existence of outlier items in rankings and their impact on click data. Notice that we use this data only for click analysis and parameter estimation (Section~\ref{sec:experiments}).

\header{Data collection}
We collect search query logs from 20 consecutive days.
Each row of the dataset consists of seven observable item features that are explained in Table~\ref{table:features}, along with users' interaction signals: impressions and clicks.

\begin{definition}[Impression]
An \emph{impression} indicates how many times an item that is rendered by the search engine is viewed by a user. If an item is rendered in low positions, it may not end up in a window that is visible to the user, leading to zero impressions. On the other hand, the number of impressions can be greater than one due to scrolling.
\end{definition}

\noindent%
We selected item features that are used across different categories, are observable by users, and have been shown
by previous work to be important in influencing users' purchase decisions~\citep{aggarwal2016font, kao2020effects}.
We leave out item images from our click exploration due to the excessive complexity they would have added to this study.

Most search engines consider diversity as a quality of search result pages~\citep{agrawal-2009-diversifying}. This can have a side effect, where the returned rankings may contain outlier items. 
Hence, query logs are a valuable source for studying the outliers' effect on users' clicking behavior. To begin, we define two types of rankings based on the existence of outliers as follow:

\begin{definition}[Normal rankings]
We call rankings that contain no outlier \emph{\noOutlier} rankings. Normal rankings can either consist of a homogeneous set of items or a diverse set. 
\end{definition}

\begin{definition}[Abnormal rankings]
We define \emph{\withOutlier} rankings to be lists that contain at least one outlier. 
\end{definition}

\begin{table*}[t]
   \caption{Description of the observable features used to represent the items.}
   \label{table:features}
   \centering
   \begin{tabular}{lll}
     \toprule
      Feature Name  & Abbreviation &  Description \\
      \midrule
      \price & - &  Selling price of an item.\\
      promotion tag & \promotion & Universal red tag indicating various promotions, such as `competitive price' and `select deal'.\\
      high discount tag & \discount & Two-piece red tag indicating high discount for an item (different from promotion).\\
      in/out-of-stock tag & \stockQuantity & Green tag indicating the in-stock or out-of-stock condition of an item.\\
      users star rating & \rating & Average user star rating of the item presented by the standard 5 stars template.\\
      `select' tag & \select & Green tag indicating that the item is a select item (similar to Amazon prime). \\
      \titleLength & - & Number of tokens in the item title.\\
      \bottomrule
   \end{tabular}
\end{table*}

\header{Outlier detection}
We examine each item for outlierness based on the features described in Table~\ref{table:features} and in the context of all items in the list as described in Section~\ref{sec:background}. An item is an outlier if it is detected as an outlier w.r.t.~at least one of these features.

We use the Interquartile rule to detect the outliers, and consider the absolute difference between the feature value and the upper/lower bound as the degree of outlierness of the corresponding item (see Section~\ref{sec:background}). 
Feature values are normalized so that we have an outlierness degree of unit range for all observable features. We set the threshold for the degree of outlierness to 0.5, which means we only label an item as an outlier if the absolute difference between its score and upper/lower bound is greater than 0.5.

\header{Post-processing}
We filter out the parts of the rankings that are not viewed by the user based on the impression signal in our data. This leaves us with the minimum ranking size of 3. However, since by definition outlierness is meaningless in lists shorter than 4, we removed these rankings from our dataset. We also removed pages with sponsored items to avoid any potential effect from such items on our results.
The remaining 10,903 \withOutlier rankings have an average length of 10.24 and a median of 8.0.

\header{Effect of outliers on \ac{CTR}}
In the first step of our analysis, we aim to see if users interact differently with outlier items in \withOutlier rankings. 
To this end, we look at such rankings and compare 
the number of interactions outliers received on average to non-outlier items in the same ranking. We focus on clicks as interactions. 

Since \noOutlier rankings carry no information for our current analysis we only keep \withOutlier rankings.
Table~\ref{table:outliers-ctr} shows the average clicks, impressions, and \ac{CTR} of outlier and non-outlier items for \withOutlier rankings.\footnote{Note that the reported values in this section are calculated based on filtered subsets of search logs, therefore, they are not representative of the true statistics of the data.} 
We calculate the CTR values (i.e., the number of clicks divided by the number of impressions of each item) per page and report the average over all rankings. 
Our findings suggest that both \ac{CTR} and average clicks are significantly higher for outlier items when compared to non-outlier items on the same page. Moreover, we see that the number of impressions is also significantly higher for outlier items, which is in line with the finding of an eye-tracking experiment reported in~\citep{sarvi2022understanding}.

\begin{table}
    \caption{Users' interactions with the outlier and non-outlier items, averaged over all \withOutlier rankings. We used Student's t-test with $p < 0.001$ for statistical significance test.}
    \label{table:outliers-ctr}
    \centering
      \begin{tabular}{lccc}
        \toprule
        & Avg. clicks & Avg. impressions & Avg. CTR\\
      \midrule
      Outliers & 0.202\rlap{$^\ast$} &  1.381\rlap{$^\ast$} & 0.142\rlap{$^\ast$} \\
      Non-outliers & 0.137 &  1.346 & 0.098\\
      Total & 0.149 & 1.352 & 0.106\\
      \bottomrule
    \end{tabular}
\end{table}

\begin{figure}
    \includegraphics[width=\linewidth]{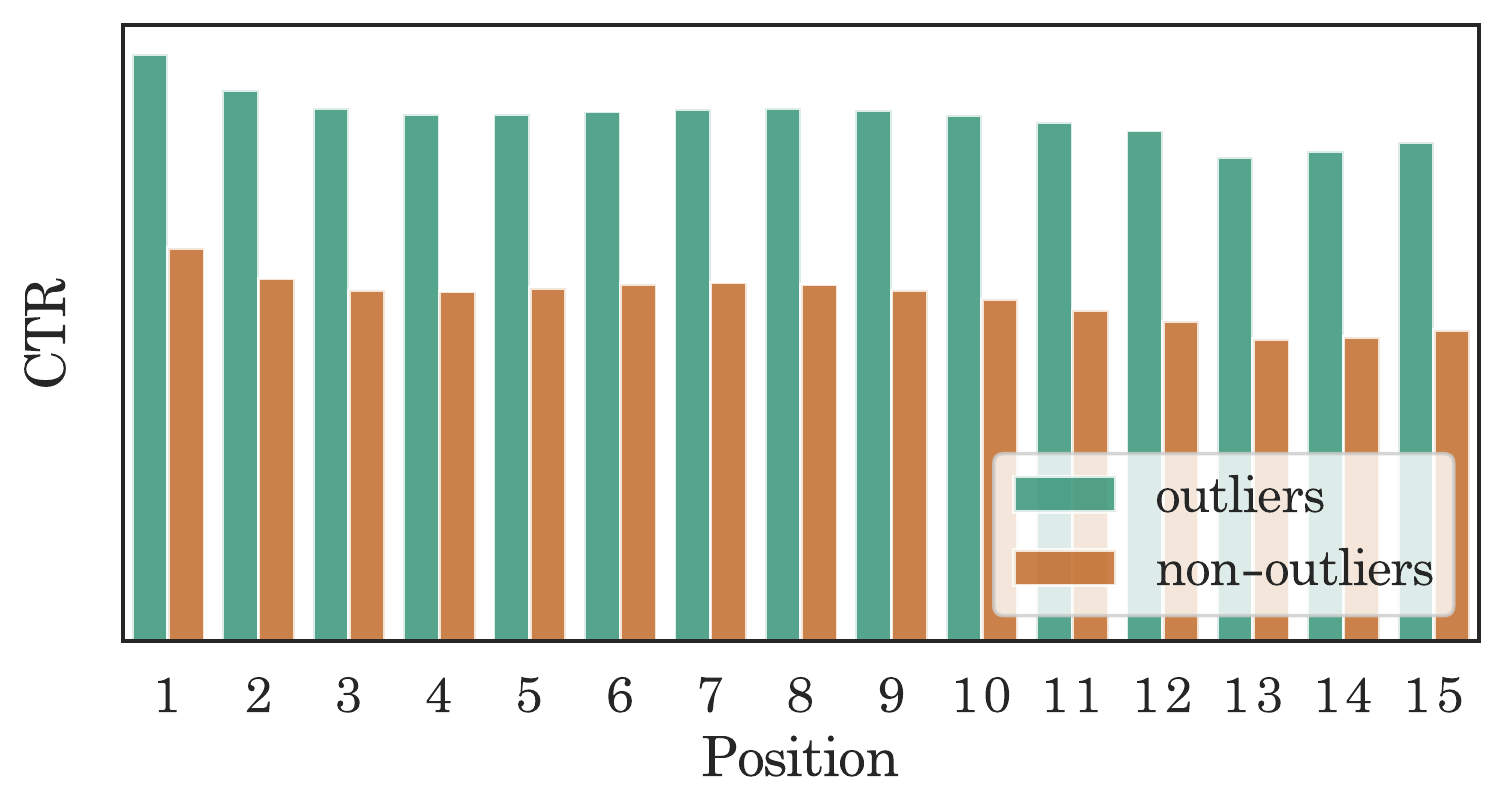}
    \caption{Comparison of \ac{CTR} for outlier and non-outlier items per rank. \ac{CTR} is consistently higher for outlier items.}
\label{fig:ctr-per-position}
\end{figure}

\header{Effect of outliers per position}
To make sure that the higher \ac{CTR} reported in Table~\ref{table:outliers-ctr} is not caused by position bias, we look at \ac{CTR} values per position. Figure~\ref{fig:ctr-per-position} depicts the results. Overall, \ac{CTR} for all positions is higher for outlier items, showing that these items receive more interactions than non-outlier items. 

\begin{figure}
    \includegraphics[width=\linewidth]{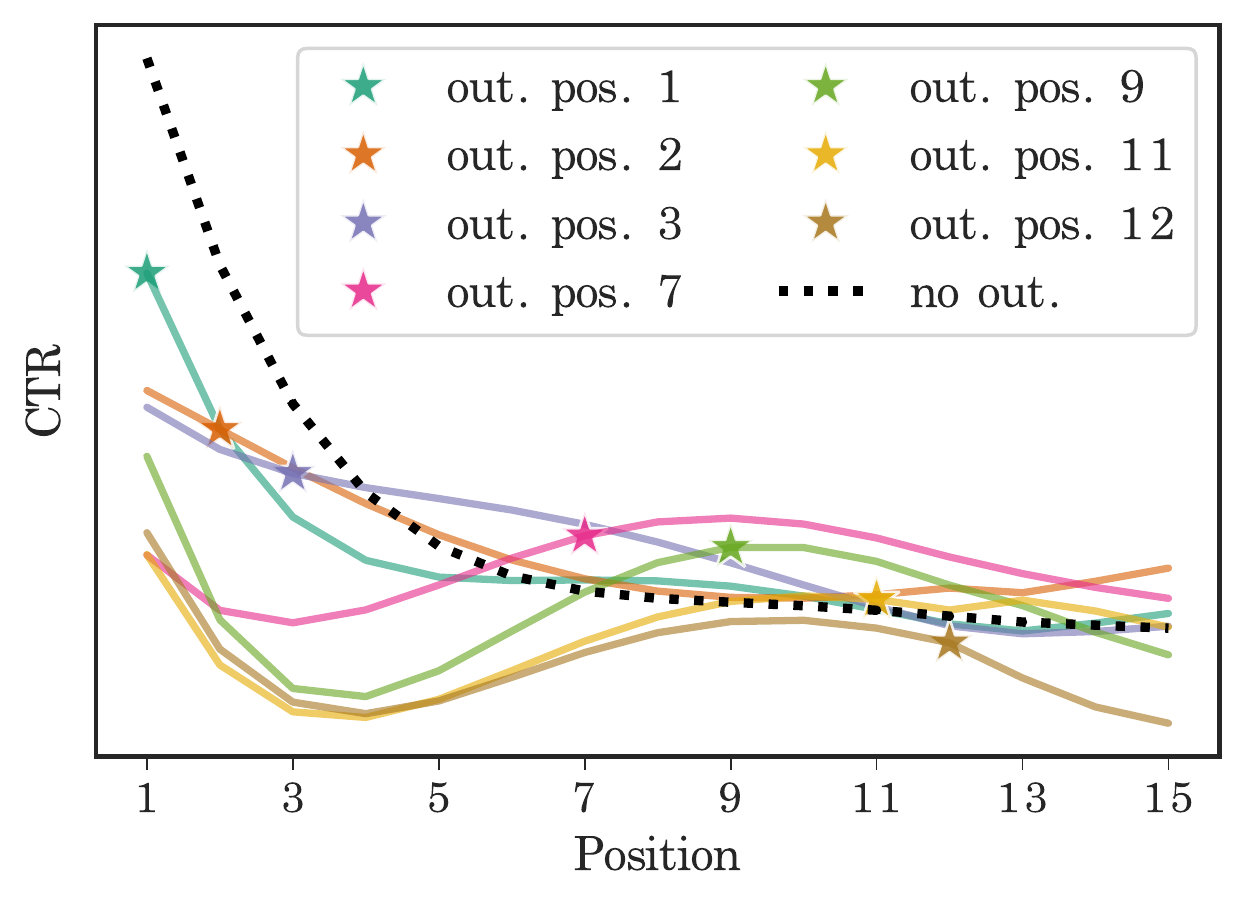}
\caption{\ac{CTR} per rank for \withOutlier rankings grouped by the outliers' position. The position of the outlier is marked with an asterisk. The values are smoothed using a Savitzky-Golay filter. Best viewed in color.\vspace{-0.1cm}}
\label{fig:ctr-outliers-per-rank}
\end{figure}

Next, to further study how outliers change users' click behavior, we compare the \ac{CTR} of the outlier position with the positions of non-outlier items throughout the ranking. 
To better depict the effect of outliers on different positions, we consider rankings that contain exactly one outlier; we focus on the top 15 positions. It is worth mentioning that less than $35\%$ of the \withOutlier rankings in our data have more than one outlier.
We group the \withOutlier rankings based on the position of the outlier. 
Figure~\ref{fig:ctr-outliers-per-rank} illustrates the results. The black line shows \ac{CTR} for \noOutlier rankings. As expected this line follows position bias, where the probability of clicking an item decreases with its rank. 

The other lines in Figure~\ref{fig:ctr-outliers-per-rank} show the \ac{CTR} for groups of rankings with one outlier at position $r\in\{1, \dots, 15\}$. We only show the results for some of the positions for better visibility. We see similar patterns for other ranks.
We only show the results for groups that form at least $1\%$ of the whole collection in terms of size. In Figure~\ref{fig:ctr-outliers-per-rank} each asterisk indicates the position of the outlier. 
We observe that \ac{CTR} distribution is different than the position-based assumption when there is an outlier in the ranking. Also, for positions after 3, we observe an increase of \ac{CTR} on and around the outlier position.

Another interesting observation is that items farther away from the outlier receive less attention proportional to their ranks compared to \noOutlier pages.
Moreover, we see that for positions after 3 \ac{CTR} for outliers is higher than the \ac{CTR} for the same position in \noOutlier pages, which is in line with our findings in Figure~\ref{fig:ctr-per-position}.

\header{Effect of different outlier types}
One can argue that different types of outliers might have different types of influence on users' perceptions. E.g., considering \price as the observable feature, a very expensive outlier item might have a lower chance of being purchased compared to a cheap one. We hypothesize that these two types may neutralize each other overall in terms of statistical metrics. Hence, to examine this hypothesis, as a first attempt, we divide the outlier items into two groups of positive and negative outliers using common sense, informal definitions based on the observable features. 
E.g., in the previous example, the expensive item is a negative outlier while the cheap one is positive. 
Based on this definition, for the \price feature we see that the average number of clicks for positive and negative outliers are 0.193 and 0.147, respectively; both are significantly higher than non-outlier items (0.125). We see the same trend among all observable features, both for impression and click counts. 
Based on these results we reject the aforementioned hypothesis and stay with our original outlier/non-outlier division.  

\header{Further remarks}
We also looked at \withOutlier rankings in which a specific item is repeatedly shown in a fixed rank at least once as an outlier and once as a normal item. We aggregate all such rankings and observe that on average items receive $0.169$ clicks in case of being an outlier, and $0.130$ clicks when they are regular items in the list.
Comparing the \withOutlier rankings to a subset of \noOutlier rankings with a similar length distribution (mean=10.09/10.30, median=8.0/8.0, std=4.95/5.82 for \noOutlier/\withOutlier rankings), we realize that on average the number of clicks per session is higher in the presence of outliers. More specifically, the average number of clicks is $0.139$ for \noOutlier rankings, and $0.149$ for \withOutlier rankings, with a $p<0.001$ significance.   

\header{Upshot}
To sum up, from Section~\ref{sec:outlier-impact:user-study} we learn that users behave differently w.r.t.\ an item given its outlierness condition (i.e.,~whether the item is presented as an outlier in the ranking). The \ac{CTR} of a specific item is consistently
higher when it is presented as an outlier item than that of a non-outlier item.
Section~\ref{sec:outlier-impact:logs} confirms
the findings of our user study. In addition, we observe that, on average, outlier items receive significantly more clicks than non-outlier items on the same lists. Moreover, users tend to interact more with lists that contain at least one outlier.
This section confirms the impact from  outlier items on clicks. We refer to this effect as outlier bias.
In the following section we propose a click model that accounts for outlier bias as well as position bias.

\section{\hspace{-0.95em} Outlier-aware Position-Based Model}
\label{sec:method}

Na\"ive use of implicit feedback for learning to rank can be misleading, since it suffers from presentation bias. Therefore, modeling the examination bias is crucial~\citep{joachims2007evaluating, fang2019intervention}.

\header{\Acl{PBM}}
Normally, items in higher ranks are more likely to be examined on a page. 
Position bias is formally modeled through the examination hypothesis which states that an item must be examined and perceived relevant by the user to be clicked.
A widely used click model for dealing with position bias is the \ac{PBM}~\citep{joachims2017unbiased, wang2018position}. While being considered a simple solution, \ac{PBM} is as effective as more sophisticated click models~\citep{chuklin-2015-click}.
\ac{PBM} assumes that the rank of an item is the only parameter that affects users' examination of that item.
Examining an item means viewing and evaluating it before any subsequent interaction like a click. 

Given an item $d$ at rank $k$ in response to a query $q$, the probability of clicking on $d$, assuming \ac{PBM}, equals:
\begin{equation}
    P(C=1\mid q, d, k) = P(E=1\mid k) \times P(R=1\mid d, q),
\end{equation}
where $P(E=1\mid k)$ is the probability of user examining rank $k$, also called \emph{propensity}, and $P(R=1\mid d, q)$ is the probability of relevance for the pair $(d, q)$. 
We refer to these probabilities as $\theta_k$ and $\gamma_{q, d}$, respectively. 

\header{Outlier-aware position-based model}
\ac{PBM} simply assumes that the only factor influencing the propensity is the rank. 
In Section~\ref{sec:outlier-impact} we show that users are more likely to click on outlier items, hence, we assume that propensity depends also on the existence of outlier item(s).

It is noteworthy that, even among the outlier items we observe an inter-outlier position bias -- the higher-ranked outlier items receive more clicks. 

Hence, to model these dependencies, we propose an outlier-aware position-based model, called \ac{OPBM}, that accounts for the impact of outlier items in addition to the position as follows: 
\begin{equation}
\label{eq:opbm}
    P(C=1\mid q, d, k, o) = P(E=1\mid k, o) \times P(R=1\mid d, q),
\end{equation}
where $o$ indicates the position(s) of the outlier(s) in the ranking. 
Note that \ac{PBM} is a special case of \ac{OPBM}: for \noOutlier rankings \ac{OPBM} is simplified to \ac{PBM}.

We propose this model following Eq.~\eqref{eq:opbm} based on the assumption that the probability of examination at rank $k$ depends on the position of outlier item(s), $o$, in addition to $k$.
This model has $K\times O$ parameters, where $K$ and $O$ are the set of all ranks and outlier positions, respectively, which can be estimated from click data. 

\header{Propensity estimation}
Here, we describe how to estimate outlier-aware position bias from regular clicks.
Based on the idea of the \ac{REM} algorithm~\citep{wang2018position}, we propose to estimate the parameters $\theta_{k,o}$ and $\gamma_{q,d}$ simultaneously by estimating with a regression function.

Using a \ac{EM} algorithm we aim to find the parameters that maximize the log-likelihood of the whole click logs.
The log likelihood of generating click logs of the form $\mathcal{L}=(c, q, d, k,o)$ is:
\begin{equation}
\label{eq:likelihood}
\mbox{}\hspace*{-2mm}
    \log P(\mathcal{L}) = \sum_{(c, q, d, k,o)\in \mathcal{L}} c\log \theta_{k,o}\gamma_{q, d} + (1-c)\log (1 - \theta_{k, o}\gamma_{q, d}).
\end{equation}
Here, we aim to estimate the parameters $\theta_{k,o}$ and $\gamma_{q, d}$ based on data points in $\mathcal{L}$. In each iteration, \ac{EM} alternates between the expectation and maximization steps to compute new estimates of the parameters. 
In the expectation step of iteration $t+1$ we calculate the hidden variables corresponding to examination propensity ($E$) and true relevance ($R$) based on the estimated parameters at iteration $t$:
\begin{align}
  &P(E = 1,R = 1\mid C = 1,q,d,k,o) = 1,
  \nonumber\\
  &P(E=1,R=0\mid C=0,q,d,k,o) = \frac{\theta_{k,o}^t(1-\gamma^t_{q,d})}{1-\theta_{k,o}^t\gamma^t_{q,d}},
  \nonumber\\
  &P(E=0,R=1\mid C=0,q,d,k,o) = \frac{(1-\theta_{k,o}^t)\gamma^t_{q,d}}{1-\theta_{k,o}^t\gamma^t_{q,d}},
  \label{eq:expectation-step}\\
  &P(E=0,R=0\mid C=0,q,d,k,o) = \frac{(1-\theta_{k,o}^t)(1-\gamma^t_{q,d})}{1-\theta_{k,o}^t\gamma^t_{q,d}}.
  \nonumber
\end{align}
We then calculate the marginal probabilities $ P(E = 1\mid c,q,d,k,o)$ and $P(R = 1\mid c,q,d,k)$ for each data point in $\mathcal{L}$. 
We keep the estimation of $\gamma_{q,d}$ untouched, meaning that the \ac{LTR} model is trained without  knowledge of the outlier position and only the propensity estimation is affected by that. 
This leads to the maximization step at iteration $t+1$, where we update the parameters to maximize the likelihood from Eq.~\ref{eq:likelihood} as follows:
\begin{align}
    &\theta^{t+1}_{k,o}=\frac{\sum_{c, q, d, k',o'}\mathds{I}_{k'=k,o'=o}. (c+(1-c)P(E=1\mid c,q,d,k,o))}{\sum_{c, q, d, k',o'}\mathds{I}_{k'=k,o'=o}},
    \nonumber\\[1mm]
    &\gamma^{t+1}_{q,d} =\frac{\sum_{c, q', d', k}\mathds{I}_{q'=q,d'=d}. (c+(1-c)P(R=1\mid c,q,d,k))}{\sum_{c, q', d', k}\mathds{I}_{q'=q,d'=d}}.
\label{eq:maximization-step}
\end{align}
The maximization step of the \ac{EM} algorithm requires multiple occurrences of pair $(q, d)$ where $d$ is shown in different positions. To overcome the click sparsity problem and possible privacy issues, we alter the maximization step at iteration $t+1$, 
 where we estimate the $\gamma_{q,d}$ parameter via regression~\citep{wang2018position}. Thus, given a feature vector $x_{q,d}$ representing the pair $(q,d)$ we fit a function $f(x_{q,d})$ (e.g., \ac{GBDT}) to calculate an estimate for $\gamma_{q,d}$. So, our maximization step is to find a regression function $f(x)$ that maximizes Eq.~\ref{eq:likelihood} given the estimated parameters from the expectation step. In \ac{REM} algorithm~\citep{wang2018position}, this regression problem is  converted to a classification problem by sampling a binary variable indicating the relevance label for $x_{q,d}$ from the distribution $P(R=1\mid c,q,d,k)$. This results in a training set of the form ${(x_{q,d}, r_{q,d})}$ with the following \acl{CE} objective:
\begin{equation}
\label{eq:bce}
    \sum_{x,r}r\log (f(x)) + (1 - r)\log (1 - f(x)).
\end{equation}
 
\begin{remark}
An alternative choice instead of a single unbiased model would be to train multiple \ac{LTR} models as unbiased experts for different outlier positions. This alternative has two main drawbacks. First, having experts means that each expert is trained only on a part of the data containing outliers at a specific position. Not only can this lead to sub-optimal training, but it also makes it difficult to compare this model to the \ac{PBM}-based \ac{REM} as a baseline. Second, having a collection of $K$ expert models as a ranker is not ideal in real-world scenarios. Ideally, there is a single unbiased model that can be used without  information about outliers' positions.
\end{remark}

\section{Experimental Setup}
\label{sec:experiments}
Following much previous work in \ac{CLTR}~\citep{ai2018unbiased, vardasbi2020inverse, joachims2017unbiased,jagerman2019model,oosterhuis2020policy}, we use a semi-synthetic setup for our experiments, i.e., we sample queries, documents, and relevance labels from existing \ac{LTR} datasets, but simulate user clicks based on the probabilistic click models estimated on the \propdata data.

\ac{LTR} datasets that contain the true relevance labels allow us to evaluate the relevance estimation of \ourmodel and other baselines, as well as their effect on ranking performance. 
Furthermore, the semi-synthetic setup enables us to control the position bias and outlier bias of the simulated clicks. 

\vspace*{-0.5mm}
\subsection{Data}
\textbf{Public \ac{LTR} data.}
Following prior work on \ac{CLTR}~\citep{joachims2017unbiased, vardasbi2020inverse, vardasbi2021mixture}, we use the Yahoo! Webscope~\citep{Chapelle2011} and MSLR-WEB30k~\citep{qin2013introducing} datasets.
In both datasets, there are a total of around $30$k queries, each associated with a list of documents.
The query-document feature vectors of the Yahoo!\ and MSLR datasets have dimensions $501$ and $131$, respectively.
Both datasets have graded relevance labels with $5$ levels.
We follow prior work and take grades $\{3,4\}$ as relevant and grades $\{0,1,2\}$ as non-relevant.
The training sets of the Yahoo!\ and MSLR datasets have $20$k and $19$k queries with $473$k and $2.2$M documents, respectively.
The test sets of the Yahoo!\ and MSLR datasets, have $6.7$k and $6$k queries with $163$k and $749$k documents, respectively.

\header{Proprietary data}
We use the real-world click log data as described in Section~\ref{sec:outlier-impact:logs} for the experiments and refer to it as \emph{\propdata} data. We use a feature vector of size 24 containing both the relevance features and products' observable features to present each query-document pair. We use these features for the \ac{LTR} model. 
We also use the setup described in Section~\ref{sec:outlier-impact:logs} to detect the outlier items, using the Interquartile rule, w.r.t.~the observable features (see Table~\ref{table:features}). Since the rankings in this dataset have an average length of 10.24 and a median of 8.0, we use the top-$10$ items in the experiments.

\vspace*{-0.5mm}
\subsection{Click simulation}
\label{sec:experiments:click-simulation}
We follow prior work~\citep{joachims2017unbiased, vardasbi2020inverse, oosterhuis2020policy, ai2018unbiased, vardasbi2021mixture} and sample $1\%$ of the queries from each public dataset, uniformly at random, to train an artificial production ranker.
We apply probabilistic click models on rankings produced by this production ranker to simulate clicks for the semi-synthetic experiments. 
We apply our \acl{OPBM} with different approximations for examination probabilities. The relevances $\gamma_{q,d}$ are based on the relevance label recorded in the datasets. Following previous work~\citep{joachims2017unbiased, vardasbi2020inverse} we use binary relevance:
\begin{equation}
\label{eq:binary-relevance}
P(R=1\mid q,d)=\gamma_{q,d}=
    \begin{cases}
        1 & \text{if relevance\_label}(q,d) > 2\\
        0 & \text{otherwise} 
    \end{cases}.
\end{equation}
To simulate the outlier bias we follow two strategies as follows:

\header{\smash{\bolOpbm}}
We use the propensities estimated by \ourmodel (see Section~\ref{sec:method}) on our \propdata dataset. 
From all the \withOutlier rankings in our dataset, $64\%$ contain only one outlier. 
Since improving ranking for more than half of queries can lead to significant improvement in real-world scenarios, we first address this majority case. 
Therefore, with this model,
we focus on rankings with one outlier.
Thus, the output of \ourmodel is at most a $K\times K$ matrix, corresponding to all combinations of rank and outlier position, where $K=10$ in our experiments.
We use this matrix to approximate $P(E=1\mid k, o)$.

\header{\smash{\normalOpbm}}
Here, we assume that an outlier's effect on the user clicks follows a Gaussian distribution, centered at the outlier's position.
Therefore, for each $k$, we compute the linear interpolation of outlier bias, and position bias distributions, as follows:
\begin{equation}
\label{eq:normal-opbm}
   OPBM_{\mathcal{G}}(q,d,k,o) = \gamma_{q,d}((1-\alpha)\theta_k + \alpha \mathcal{G}(\mu=o,\sigma^2)),
\end{equation}
where $\mathcal{G}$ is a Gaussian distribution with $\mu=o$, simulating the outlier effect. 
We set $\sigma = 1$ and experiment with varying values of $\alpha$. 
To simulate clicks for rankings with multiple outliers, we compute the average of \normalOpbm for all outlier positions ($O^\prime$) as follows: 
\begin{equation}
\label{eq:normal-opbm-multiple-outliers}
   OPBM_{\mathcal{MG}}(q,d,k,O^\prime) = \frac{1}{|O^\prime|} \sum_{i\in O^\prime} OPBM_{\mathcal{G}}(q,d,k,i).
\end{equation}
According to our \propdata data, $91\%$ of \withOutlier rankings contain at most two outliers. Therefore, in the experiments, we focus on rankings with a maximum of two outliers.

We follow previous work~\citep{vardasbi2020inverse, jagerman2019model, joachims2017unbiased,oosterhuis2020policy} to define the position bias inversely proportional to the item's rank as:
\begin{equation}
\label{eq:experiments:position-bias}
\theta_k = \frac{1}{k}.
\end{equation}
We train the LTR model\footnote{We use allRank implementation for our LTR~\citep{Pobrotyn2020ContextAwareLT}.} on $1$M simulated clicks.

\subsection{Methods used for comparison} 
Our main goal is to introduce a new type of bias and study its impact on click propensities.
Hence, it suffices to compare our outlier-aware click model to baselines that only corrects for position bias. To this end, we compare \ourmodel with the following estimators:

\begin{itemize}[leftmargin=*]
\item \textbf{Na\"ive} is a model with no correction where each click is treated as an unbiased relevance signal.
\item \textbf{PBM} is the original \ac{IPS} estimator~\citep{joachims2017unbiased, wang2018position} that only corrects for position bias.
\end{itemize} 
 
\subsection{Evaluation metrics}
To measure the ranking performance achieved by different methods we use \ac{NDCG}.
We also consider \ac{CE}, which measures the difference between the true relevance and unbiased relevance calculated by the estimator; it is an indication of how accurately a model estimates the relevance, independent of the \ac{LTR} model. Since we work with binary relevance, we compute binary CE between the corrected clicks, i.e., $c/\theta_{k}$ and $c/\theta_{k,o}$ for PBM and OPBM, respectively, as predictions and the true relevance values as labels.
We report the mean value of CE instead of its summation, for better readability.

\begin{figure}
      \includegraphics[width=0.66\linewidth]{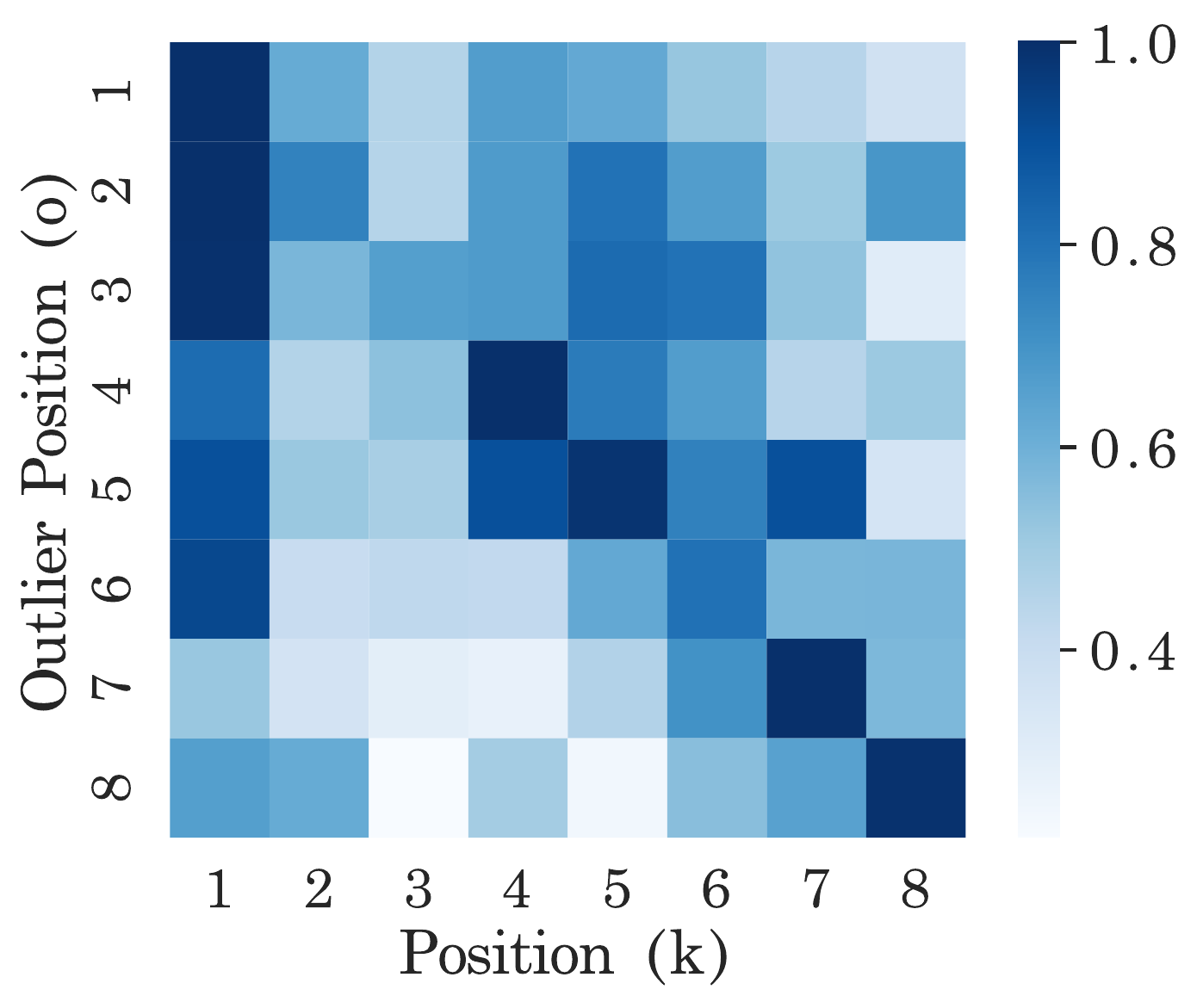}
    \caption{Click propensities computed by \ourmodel for the top 8 ranks, per outlier position on the proprietary data. Click propensities are higher on and around the outliers, contradicting the position bias assumption.}
\label{fig:results:heatmap}
\end{figure}

\section{Results}
\label{sec:results}

In Section~\ref{sec:outlier-impact} we have already answered~\ref{RQ1} about the existence of outlier bias in ranked lists.
In this section we answer the following research questions: 
\begin{enumerate*}[nosep,label=(\textbf{RQ\arabic*}),leftmargin=*]
    \setcounter{enumi}{1}
    \item how does our outlier-aware model, \ourmodel, perform compared to the baselines? \label{RQ2} 
    \item how does \ourmodel perform under different outlier bias severity conditions?\label{RQ3} 
    \item how does \ourmodel generalize to cases with multiple outliers in rankings?\label{RQ4} 
\end{enumerate*}

\begin{figure*}[ht]
    \begin{tabular}{@{}c@{}c@{}c@{}c@{}c@{}}
      \subfloat[\mslr]{%
        \includegraphics[width=.2475\linewidth]{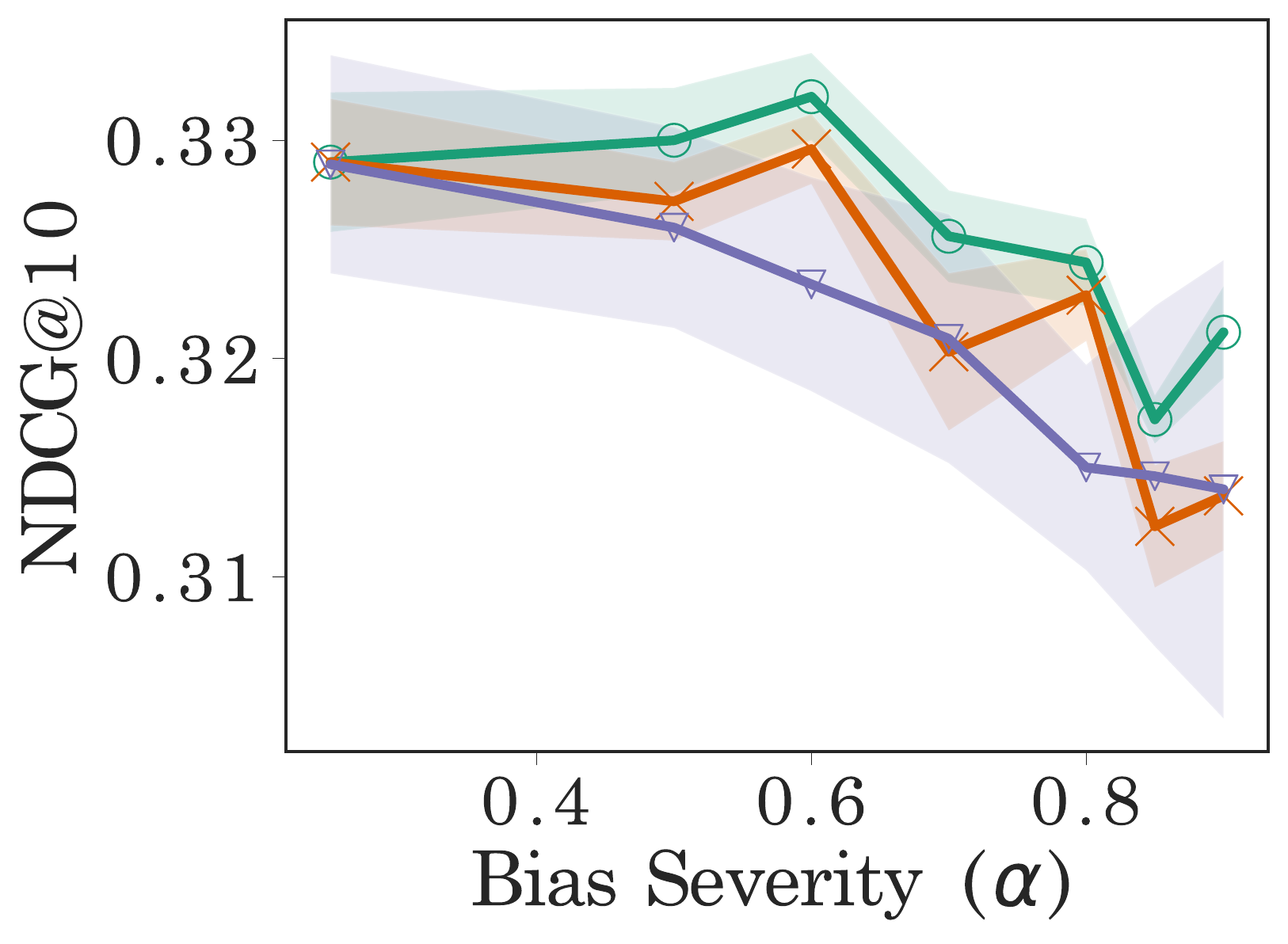}
          \label{fig:result:ndcg-mslr}}
     &
    \subfloat[\yahoo]{%
          \includegraphics[width=.2475\linewidth]{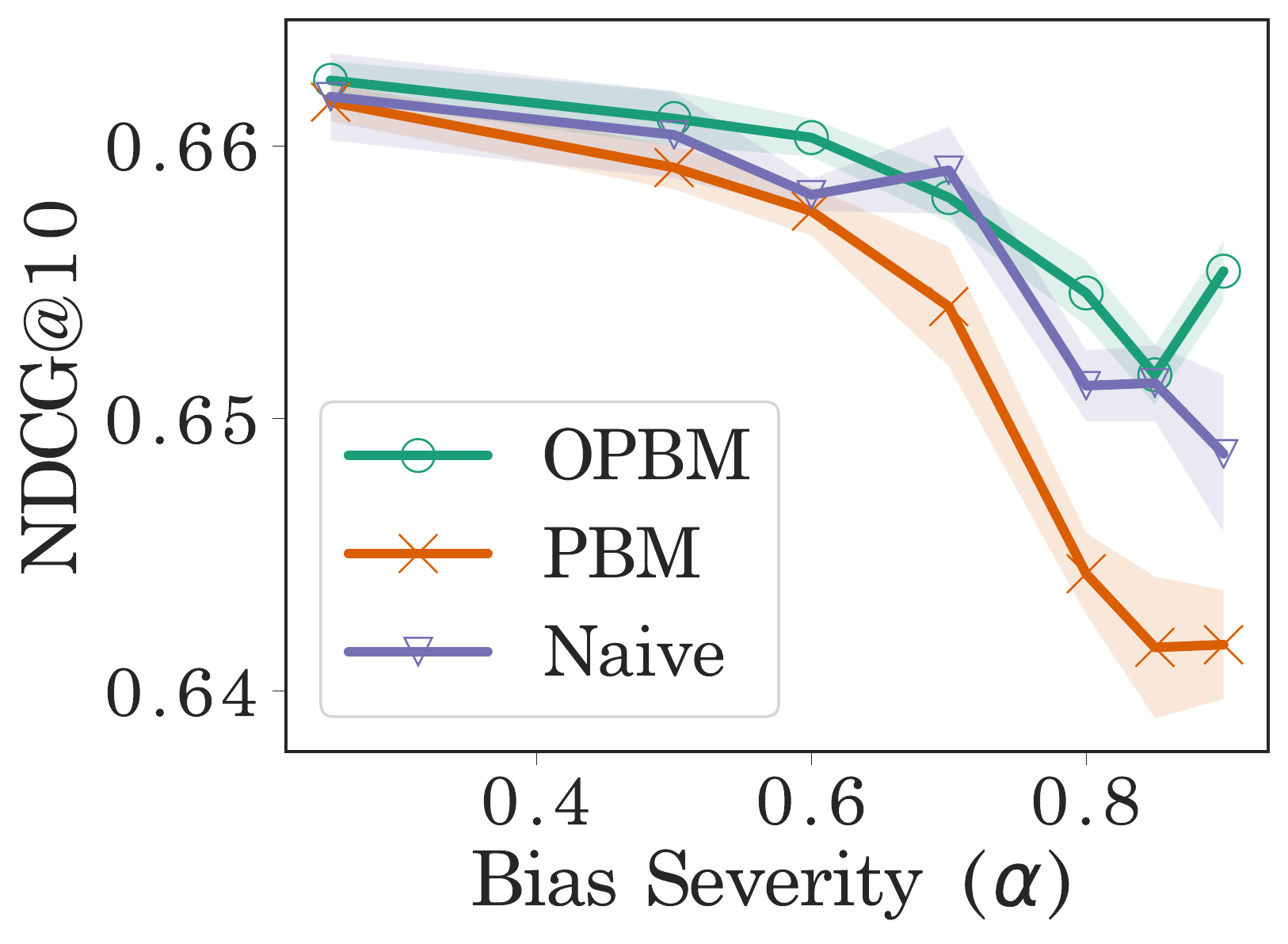}
          \label{fig:result:ndcg-yahoo}}
  &
    \subfloat[\mslr]{%
        \includegraphics[width=.2475\linewidth]{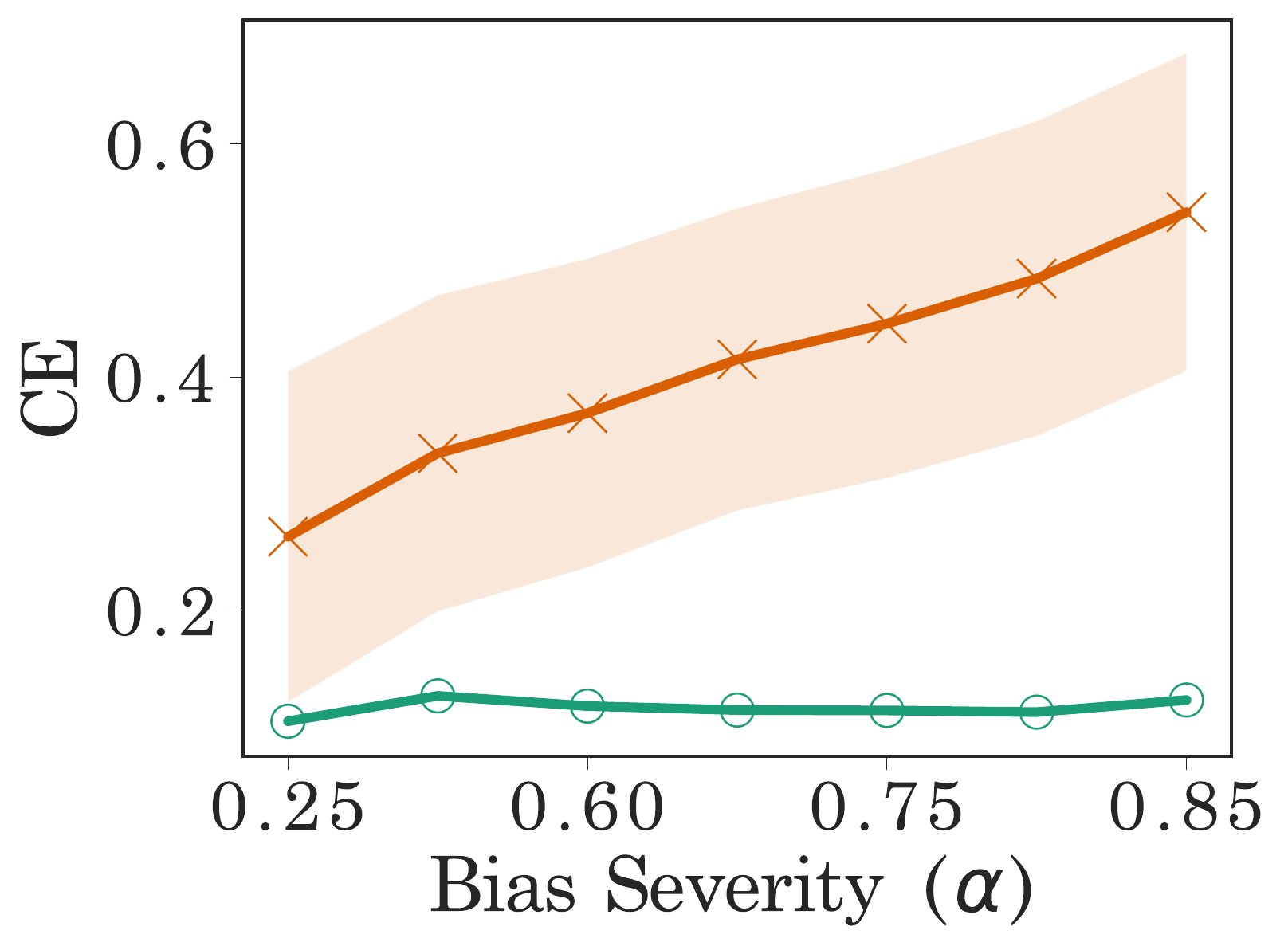}
          \label{fig:result:ce-mslr}}
     &
    \subfloat[\yahoo]{%
    \includegraphics[width=.2475\linewidth]{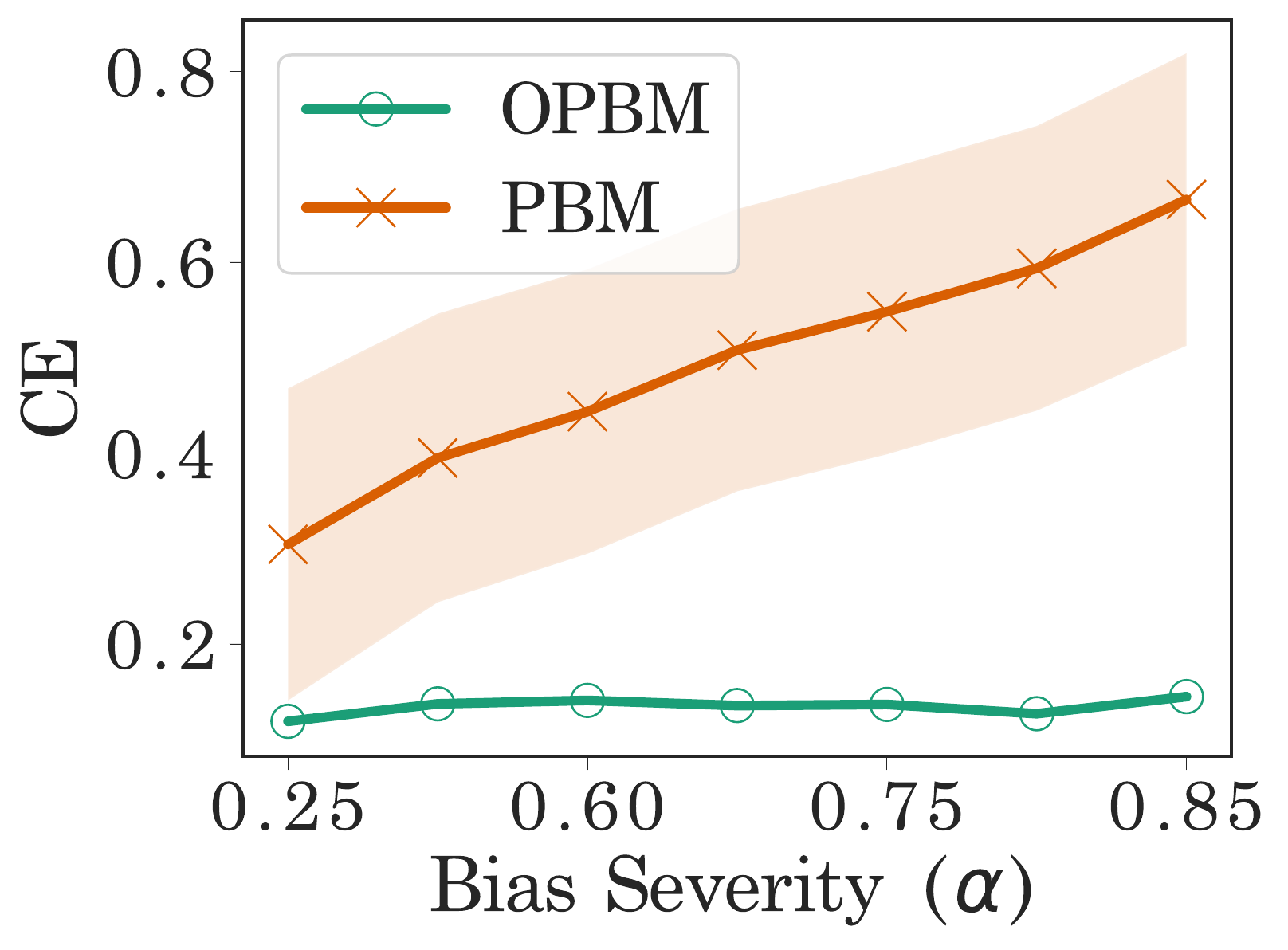}
          \label{fig:result:ce-yahoo}} 
    \end{tabular}
\caption{Comparison of different estimators in term of NDCG@10 ((a) \mslr and (b) \yahoo) and CE ((c) \mslr and (d) \yahoo) under varying levels of outlier bias. Results are averaged over 8 runs; shaded area indicates the standard deviation. 
\vspace{-0.25cm} }  
\label{fig:result:ce-ndcg}
\end{figure*}

\begin{table}
\caption{Comparison of OPBM and PBM on the \yahoo and \mslr datasets, in terms of NDCG@10 and CE. A superscript $\ast$ indicates a significant difference compared to the second-best performing method with $p<0.001$.}
\label{tabel:results-main}
\setlength{\tabcolsep}{3.1pt}
\centering
  \begin{tabular}{llcccccc}
    \toprule
    \multicolumn{1}{c}{} &&
    \multicolumn{2}{c}{\mslr} &
    \multicolumn{2}{c}{\yahoo} \\
    \cmidrule(r){3-4}
    \cmidrule(r){5-6}
 &&  
    \multicolumn{1}{c}{CE$\downarrow$} &
    \multicolumn{1}{c}{NDCG@10$\uparrow$} &
    \multicolumn{1}{c}{CE$\downarrow$} &
    \multicolumn{1}{c}{NDCG@10$\uparrow$}\\

    \midrule
    Oracle      && - &0.3451  & - &0.6713 \\
    Na\"ive     && 0.8205 &0.3065& 0.9786 &0.6489 \\
    PBM &&0.5474 &0.3165& 0.6807 &0.6406 \\
    OPBM && 0.1732\rlap{$^\ast$} & 0.3233\rlap{$^\ast$} & 0.1916\rlap{$^\ast$} & 0.6470\rlap{$^\ast$} \\
    \bottomrule
  \end{tabular}
\end{table}

\subsection{Propensity estimation with \ourmodel}
We answer \ref{RQ2} by comparing the overall performance of \ourmodel in propensity estimation. 
Figure~\ref{fig:results:heatmap} depicts the propensities estimated by \bolOpbm (see Section~\ref{sec:experiments:click-simulation}) on the top-$8$ ranks where a sufficient number of outliers exist in our proprietary dataset. We see that the propensities are highest on and around the outlier positions which is in line with our findings in Section~\ref{sec:outlier-impact}. However, this effect is less evident in the top-$3$ ranks. This is expected since we observe that position bias dominates in the top-$3$ ranks (see Section~\ref{sec:outlier-impact:logs}), diminishing the effect of outliers. Nevertheless, the effect of position bias decreases as the outlier appears higher in the ranking. For example, when an outlier occurs at position 1, the propensities of the first two ranks are 0.99 and 0.62, respectively. However, when the outlier occurs at position 7, these values reduce to 0.52 and 0.35.

Next, we report the results of the semi-synthetic experiments. We use the \mslr and \yahoo public \ac{LTR} datasets with simulated clicks. We use the propensities calculated by \bolOpbm trained on our proprietary data. 
We compare \ourmodel and \ac{PBM} in terms of relevance estimation (\ac{CE}) and ranking performance (NDCG@10). 
Table~\ref{tabel:results-main} summarizes the results; on both datasets \ourmodel performs significantly better than \ac{PBM} in terms of \ac{CE} ($p < 0.001$), indicating that \ourmodel approximates click propensities more effectively -- it estimates true relevance of a $(q,d)$ pair more accurately. Providing an accurate estimate of true relevance is crucial in domains such as exposure-based fair ranking~\citep{singh2018fairness, biega2018equity,morik2020controlling}, where relevance is used as an indication of an item's merit~\citep{singh2018fairness, biega2018equity,morik2020controlling, sarvi2022understanding, heuss2022fairness, vardasbi2022probabilistic}, and can have a big impact on fairness estimation.
Table~\ref{tabel:results-main} also shows that \ourmodel significantly improves the ranking scores (NDCG@10) over the \ac{PBM} baseline, again on both datasets.

In conclusion, using \ourmodel leads to more accurate propensity estimations and a more accurate approximation of the true relevance in rankings affected by outlier items. We also observe significant improvements in ranking performance by \ourmodel over \ac{PBM} on the \yahoo and \mslr datasets.

\begin{table}
\caption{Comparison of OPBM, OPBM$_{lazy}$ and OPM on the \yahoo and \mslr datasets, with outlier bias severity of $\alpha=0.75$, and in terms of NDCG@10 and CE. 
A superscript $\ast$ indicates a significant difference with \ac{PBM} with $p<0.001$.
}
\label{tabel:results-RQ4}
\setlength{\tabcolsep}{3.1pt}
\centering
  \begin{tabular}{llcccccc}
    \toprule
    \multicolumn{1}{c}{} &&
    \multicolumn{2}{c}{\mslr} &
    \multicolumn{2}{c}{\yahoo} \\
    \cmidrule(r){3-4}
    \cmidrule(r){5-6}
 &&  
    \multicolumn{1}{c}{CE$\downarrow$} &
    \multicolumn{1}{c}{NDCG@10$\uparrow$} &
    \multicolumn{1}{c}{CE$\downarrow$} &
    \multicolumn{1}{c}{NDCG@10$\uparrow$}\\
    \midrule
     Na\"ive     && 0.5704&0.3159 & 0.6776 &0.6564 \\
     PBM && 0.3126 &0.3219 & 0.3958 &0.6497 \\
     OPBM$_{lazy}$ &&0.1374\rlap{$^\ast$} & 0.3223& 0.1548\rlap{$^\ast$} & 0.6566\rlap{$^\ast$} \\
     OPBM && 0.1283\rlap{$^\ast$} & 0.3229 & 0.1407\rlap{$^\ast$} & 0.6572\rlap{$^\ast$} \\
    \bottomrule
  \end{tabular}
\end{table}

\subsection{Effect of outlier bias severity}
Next, we address~\ref{RQ3} by considering the impact of outlier bias severity on the performance of \ourmodel. For the sake of simplicity, we assume that outliers have the same effect on propensity distribution independent of their position; we use \normalOpbm (see Section~\ref{sec:experiments:click-simulation}) for click simulation. The parameter $\alpha$ in \normalOpbm allows us to control outlier bias severity. Figure~\ref{fig:result:ce-ndcg} depicts the results. \ourmodel consistently outperforms \ac{PBM} in terms of ranking performance.
The results on \yahoo dataset (Figure~\ref{fig:result:ndcg-yahoo}) clearly show that the difference in ranking performance of the two models increases with the severity of outlier bias. In the case of \mslr (Figure~\ref{fig:result:ndcg-mslr}) we observe more fluctuations in \ac{OPBM}'s performance. This is also visible in the high variance of Na\"ive's performance in different runs; \ourmodel performs more robust compared than Na\"ive and \ac{PBM}.
Moreover, the results show that \ourmodel performs similarly to \ac{PBM} at its worst, making it a more reliable choice as a user examination model for all severity levels of outlier bias. This is in line with our theory, which indicates that \ac{PBM} is a specific case of \ourmodel (see Section~\ref{sec:method}).

In terms of \acl{CE} (Figure~\ref{fig:result:ce-mslr} and~\ref{fig:result:ce-yahoo}), \ourmodel consistently outperforms \ac{PBM} with a high margin. Also, the high variance in performance of \ac{PBM} emphasizes the much more robust performance of \ourmodel compared to \ac{PBM} in relevance estimation.

In conclusion, using the \ourmodel estimator leads to improved ranking models compared to \ac{PBM}, especially when severe outlier bias exists. This finding also holds for accurately estimating the true relevance scores (i.e., CE). In the presence of slight outlier bias, \ourmodel exhibits a similar performance compared to \ac{PBM}, making it a natural choice as it proves to be reliable.

\begin{figure}[t]
    \begin{tabular}{@{}c@{~}c@{}}
      \subfloat[Over all rankings]{%
        \includegraphics[width=.48\linewidth]{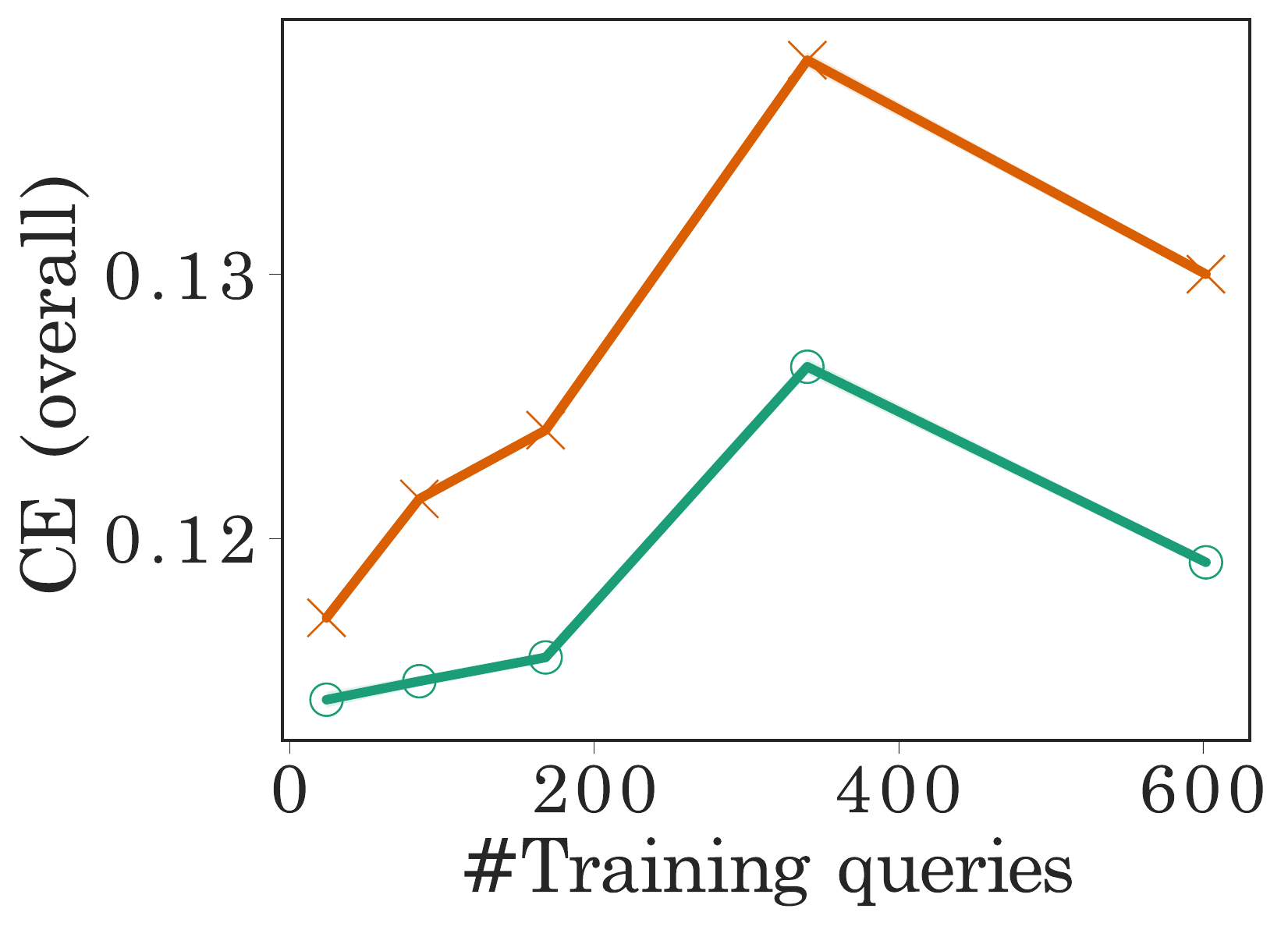}
          \label{fig:result:size-overall}}
     &
    \subfloat[Over two outlier rankings]{%
          \includegraphics[width=.495\linewidth]{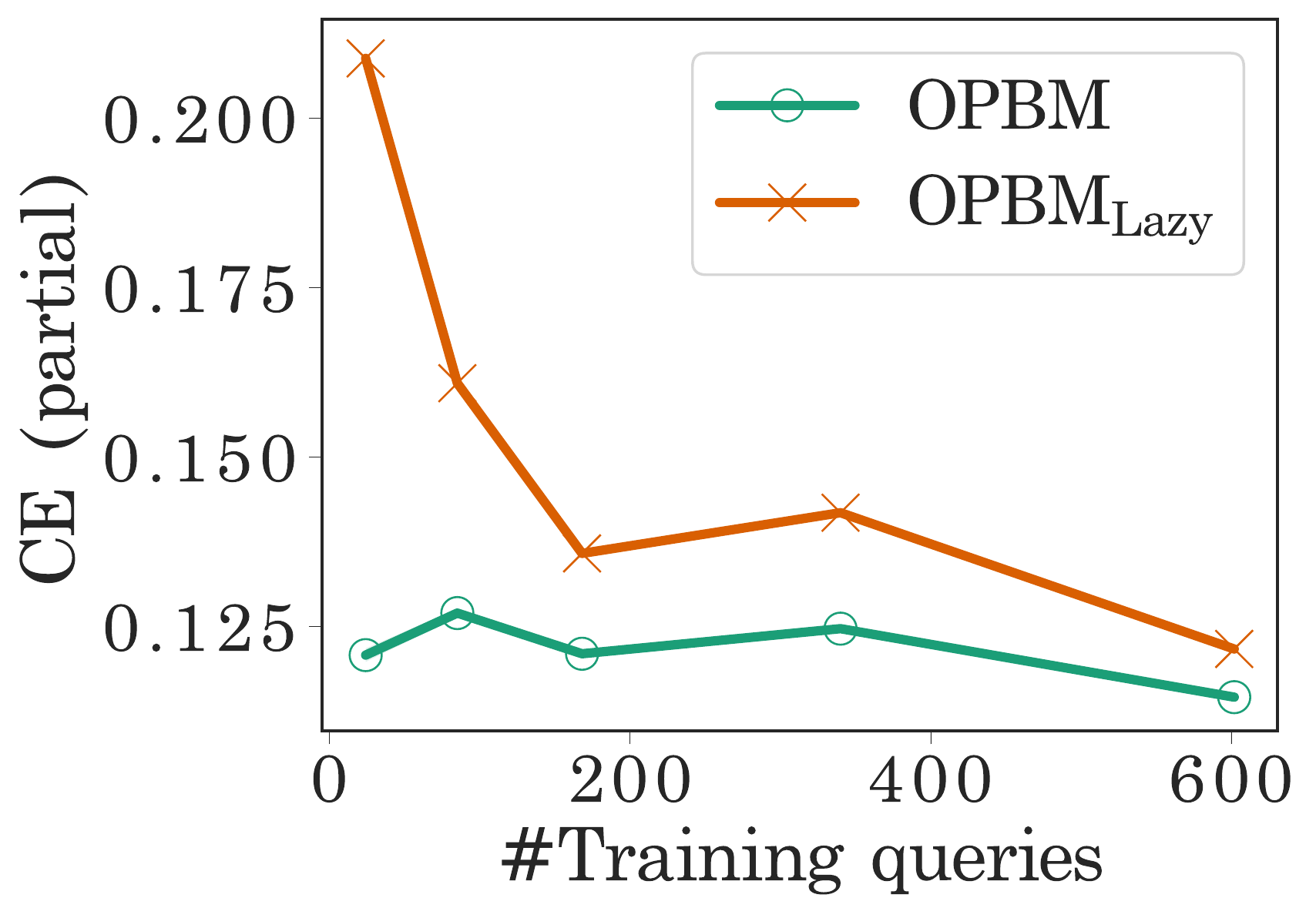}
          \label{fig:result:size-partial}}
    \end{tabular}
\caption{Comparison of \ourmodel and \ourmodelsimple on varying sizes of queries with multiple outliers.}  
\label{fig:result:size}
\end{figure}

\vspace*{-0.7mm}
\subsection{Generalization to multiple outliers}
We address~\ref{RQ4} by considering how \ourmodel generalizes to multiple outliers in the ranking. For click simulation, we use Equation~\ref{eq:normal-opbm-multiple-outliers} with severe outlier bias ($\alpha=0.75$). As pointed out before, our proprietary data shows that $91\%$ of \withOutlier rankings contain at most two outliers. Therefore, we report results for $|O^\prime|=2$. 
Here, in addition to the single outlier rankings from the previous experiments, our semi-synthetic data contains rankings with two outliers at positions 4 and 9. 
As mentioned earlier, position bias is severe in the top-$3$ ranks, thus we place the first outlier in the fourth position of the list. Then, in order to see the effect of the outliers separately, we choose the second positions with some distance (rank 9).

To model the effect of multiple outliers, we propose two strategies: 
\begin{enumerate*}[label=(\roman*)]
    \item According to the original description of \ourmodel (see Section~\ref{sec:method}), we consider the condition of having multiple outliers as a separate value for $o$, i.e., we separately compute the click propensities for $k$ ranks, when two outliers exist in the ranking at positions 4 and 9.
    \item We simplify the problem and only consider the first outlier position, and call it \ourmodelsimple.
\end{enumerate*}
We compare the performance of \ourmodel between these two strategies and also with \ac{PBM}.
Table~\ref{tabel:results-RQ4} summarizes the results. Overall, we see that both variations of \ourmodel outperform \ac{PBM} in terms of NDCG@10 and \ac{CE}. As expected, \ourmodel outperform its simpler version, \ourmodelsimple w.r.t.~both metrics; we see significant improvements in \ac{CE}, while the improvements over ranking performance are marginal. We can conclude that the original version of \ourmodel as the exact solution performs better for cases with multiple outliers. However, in case of data sparsity we can reduce the problem to the single outlier setup and still achieve higher results than \ac{PBM}.

Lastly, we provide insights into how the size of the training data influences the performance of \ourmodel compared to \ourmodelsimple. We gradually increase the number of training queries for the rankings with two outliers (positions 4 and 9), while keeping the rest of the training set unchanged. We compare the performance in term of \ac{CE} on all rankings (overall), and only on the two outlier rankings (partial). See Figure~\ref{fig:result:size}. We see that even with $24$ rankings with two outliers, \ourmodel manages to learn the propensities better than the \ourmodelsimple. However, the difference  between the total performance (Figure~\ref{fig:result:size-overall}) of the two models grows by the size of training samples for the two outliers rankings, suggesting \ourmodel as a natural choice when a reasonable amount of training data is available. 

In conclusion, when enough samples corresponding to multiple outlier positions are available in the training data, it is best to use \ourmodel with a specific $o$ that represents the case at hand. Otherwise, reducing these samples to the single outlier setting, by only considering the first outlier position, still outperforms \ac{PBM}. 


\section{Related Work}
\label{section:relatedwork}

\textbf{Outliers.}
An outlier is an exceptional object that deviates from the general data distribution~\citep{wang2019progress}. 
Outliers can affect the statistical analysis, whether they are interesting observations or suspicious anomalies.
Identifying these outlaying samples is crucial in many fields of study~\citep{wang2019progress,li2020copod}. 
Numerous approaches have been proposed to detect outliers~\citep{rousseeuw1999fast, ramaswamy2000efficient,scholkopf2001estimating,zhao2019lscp,li2020copod,wen-2006-ranking}. 
Defining and dealing with outliers is dependent on the application domain~\citep{wang2019progress}.  
We follow the definition of outliers in ranking from~\citep{sarvi2022understanding}: outliers are items that stand out in the ranking w.r.t.~observable item features. 
They study the effect of such items on the exposure distribution through eye-tracking experiments and further address the effect of outliers on exposure-based fairness. 
In contrast, in this work we focus on click bias caused by this phenomenon. 
We are the first to investigate the existence of outlier bias in real-world search click logs and to propose an \ac{ULTR} model to correct for outlier and position bias. 

\header{Bias in implicit feedback}
Users' implicit feedback, such as clicks, can be a valuable source of supervision for \ac{CLTR}~\citep{agarwal2019addressing}.
However, the bias in click data can cause the probability of a click to differ from the probability of relevance, which is misleading. 
In recent years, different types of bias have been studied, such as position~\citep{joachims2005accurately,joachims2017unbiased}, presentation~\citep{yue2010beyond}, selection~\citep{ovaisi2020correcting}, trust~\citep{agarwal2019addressing,vardasbi2020inverse}, popularity~\citep{abdollahpouri2017controlling}, and recency bias~\citep{chen2019correcting}. 
Another factor influencing the perceived relevance of items is inter-item dependency~\citep{chuklin-2015-click, sarvi2022understanding}.
We introduce outlier bias, which is a type of inter-item dependency. As outlier bias considers inter-item relationships it differs from the previously mentioned types of bias. 
Our work suggests that users tend to interact more with outlier items such that the examination probabilities assumed by position bias change when outlier items exist in the ranking.

Presentation bias~\citep{yue2010beyond} considers a related phenomenon; items with bold keywords in their titles appear more attractive. This differs from outlier bias by defining attractiveness of an item independent of its surrounding items. Moreover, adding more images to the top positions in a search result page can influence \ac{CTR}~\citep{metrikov2014whole}. However, the effect of such manipulations on click bias has not been studied.
The closest concept to our work is context bias in news-feed recommendation~\citep{wu2021unbiased}; \ac{CTR} is lower for products when surrounded by at least one very similar product than when surrounded by non-similar products. 
This differs from outlier bias, which emphasizes the difference between
the outlier and the other items. Also, observability is a key
factor in detecting outliers~\citep{sarvi2022understanding}, but context bias does not consider this factor.
Unlike previous work, we focus on the effect of outliers on clicks, which is observable by users and comes from inter-item dependencies.

\header{Unbiased learning to rank}
Unbiased learning to rank approaches train an unbiased ranking model directly with biased user feedback~\citep{ai2021unbiased}. These approaches can be classified into \acl{CLTR} algorithms~\citep{ai2018unbiased, joachims2017unbiased, wang2016learning} and the bandit learning algorithm~\citep{oosterhuis2018differentiable,wang2018efficient,yue2009interactively}. In this paper we are concerned with \ac{CLTR}. 
The key factor in \ac{CLTR} algorithms is first estimating examination probabilities~\citep{ai2018unbiased,wang2018position} and then using \ac{IPS}~\citep{joachims2017unbiased,wang2016learning} to debias  clicks. 
The estimations can be derived from online result randomization~\citep{wang2016learning}, online interleaving~\citep{joachims2017unbiased}, or intervention data harvested from multiple rankers~\citep{agarwal2019estimating}.
However,interventions can hurt user experience; \citet{ai2018unbiased} propose a dual learning algorithm to automatically learn both ranking models and propensities from offline data. 
Similarly, \citet{wang2018position} use \acl{REM} to compute the likelihood of observed clicks for each query. 
We build on~\citep{wang2018position} and propose an unbiased ranking model that corrects for both position bias and outlier bias by adding a parameter that accounts for the position of outlier(s). 


\section{Conclusion}
\label{section:conclusion}
We have introduced and studied a new type of click bias, that is, outlier bias. 
We conduct a user study to compare the \ac{CTR} for specific items in two conditions: once shown as outliers and once as non-outlier items in the list. 
We find that the \ac{CTR} is consistently higher when the item is presented as an outlier than when it is a non-outlier item.
Moreover, our analysis on real-world search logs confirms the findings of our user study. On average, outlier items receive significantly more clicks than non-outlier items in the same lists.\looseness=-1

To account for this effect, we propose \ourmodel, a click model based on the examination hypothesis, which accounts for both outlier and position bias.
We use \acl{REM} to estimate the click propensities based on our proposed click model, \ac{OPBM}.
Our experiments show (i) the superiority of \ac{OPBM} against compared models in terms of ranking performance, and (ii) that true relevance estimation outlier bias exists. We show that \ourmodel performs more robustly on all levels of outlier bias severity compared to \ac{PBM}. Moreover, our results show that \ourmodel performs similarly to \ac{PBM} in the worst case, making it a more reliable choice.

One limitation of our work is that for rankings with multiple outliers, we assume that the effect of each outlier is independent of its position and other outliers. We plan to investigate how multiple outliers on the same ranking affect each other and their surrounding items.
Finally, a natural extension of our work is to study how outlier bias can compensate for position bias in the top-$k$ ranks, and explore its use in different domains such as fairness of exposure.


\header{Data and code}
To facilitate reproducibility of our work, all code and parameters are shared at \url{https://github.com/arezooSarvi/outlierbias/}.

\vspace*{-1mm}
\begin{acks}
This research was supported by Ahold Delhaize
and the Hybrid Intelligence Center,  a 10-year program funded by the Dutch Ministry of Education, Culture and Science through 
the Netherlands Organisation for Scientific Research, \url{https://hybrid-intelligence-centre.nl}.
All content represents the opinion of the authors, which is not necessarily shared or endorsed by their respective employers and/or sponsors.
\end{acks}

\clearpage
\bibliographystyle{ACM-Reference-Format}
\balance
\bibliography{references}
\end{document}